\documentclass{aa}
\usepackage{graphicx}
\begin{document}
\title{On the secondary star of the cataclysmic variable
1RXS\,J094432.1+035738}

\author{R.E.\ Mennickent \inst{1}\fnmsep\thanks{Based on observations
obtained at the European Southern Observatory, ESO proposal 65.H-0410(A)}
\and  G.\ Tovmassian \inst{2} \and  S.V.\ Zharikov \inst{2} \and C.\
Tappert \inst{1} \and \\  J.\ Greiner \inst{3} \and B.T.\ G\"ansicke \inst{4}
\and R.E.\ Fried \inst{5}}

\institute{Dpto. de Fisica, Fac. de Cs. Fisicas y Mat.,
Universidad de Concepcion, Casilla 160-C, Concepcion, Chile\\
email: rmennick@stars.cfm.udec.cl  \and Observatorio Astronomico Nacional,
Instituto de Astronomia, UNAM, 22860 Ensenada, Mexico \and Astrophysical
Institute Potsdam, An der Sternwarte 16, 14482 Potsdam, Germany \and
Universitats-Sternwarte, Geismarlandstr. 11, 37083, Gottingen, Germany
\and Braeside Observatory, PO Box 906, Flagstaff, AZ 86002, USA}

   \offprints{R.E.\ Mennickent}
   \mail{rmennick@stars.cfm.udec.cl}
   \date{}

\abstract{
We present $V$ and $R_{c}$-band photometry and optical near-infrared
spectroscopy of the cataclysmic variable \object{1RXS\,J094432.1+035738}.
We detected
features of a cool secondary star, which can be modeled with a
red dwarf of spectral type M2$^{+0.5}_{-1.0}$ V at
a distance of 433 $\pm$ 100 pc. }

     \maketitle

\section{Introduction}
 
The optical
counterpart of the ROSAT source  \object{1RXS\,J094432.1+035738} was
identified as a cataclysmic variable (CV) in the course of follow-up
observations of CV candidates from the
Hamburg Quasar Survey (Jiang et al.\, 2000). These authors
presented a spectrum showing Balmer and \ion{He}{I} emission, typical of a
cataclysmic variable star (Warner 1995).  Two outbursts have been observed
and recorded by $VSNET$ observers in January 2001 and June 2001
(http://www.kusastro.kyoto-u.ac.jp/vsnet/), suggesting a dwarf nova nature
for the variable.
While writing this paper, we were informed that
Thorstensen \& Fenton found an orbital period of 0.1492 $\pm$ 0.0004 d for
this object (vsnet-alert 5988, Thorstensen, private communication). Here
we present a detailed analysis of the  spectrum of this target, including
the first detection of the secondary star, the analysis of the radial
velocity of the H$\alpha$ emission line and a refinement of the orbital
period.

\section{Observations and data reduction}

\subsection{Optical photometry}

We obtained
differential photometry of  \object{1RXS\,J094432.1+035738}
at Braeside Observatory, Arizona, using a 41\,cm reflector equipped with a
SITe\,512 CCD camera, during the nights of March 11 and 12, 2000 (UT). We
also obtained  $R_{c}$-band   time-resolved  photometry during two nights
in April 17-18, 2001 (UT) at the 1.5m telescope of the Observatorio
Astronomico Nacional de San Pedro Martir (OAN SPM), Baja California,
Mexico.  This telescope was equipped with a $1024\times 1024$\, pixel SITe
CCD.
The images were
corrected for bias and flat field. For the observations of April, 2001,
aperture differential photometry was carried out using the comparison star
located at $\alpha_{2000}$ = 09:44:27.25 and $\delta_{2000}$ = 03:58:09.9
and the check star located at $\alpha_{2000}$ = 09:44:26.45 and
$\delta_{2000}$ = 03:57:42.7.  An estimate  of the uncertainty  of  the
CCD photometry  was  obtained from the standard deviation  of  the
differential light curve between comparison and check star, viz.\, 0.02
mag. Since comparison and check star are slightly fainter than the
variable, the photometric error in this case is dominated by the fainter
star (i.e.\, the check star) and the above figure is an upper limit for
the uncertainty of the variable minus comparison light curve (e.g.\,
Howell et al.\, 1988). We did not obtain an absolute calibration for our
photometric dataset obtained at San Pedro Martir. However, for the
observations carried out at Braeside Observatory, we obtained $V$
magnitudes of \object{1RXS\,J094432.1+035738}  relative to the $V= 11.82$
comparison star GSC0023900958.
 
\subsection{Spectrophotometric observations}
 
The spectrophotometric observations were conducted at the
2.12\,m telescope of the Observatorio Astronomico Nacional, San Pedro
Martir, Mexico, during April 12, 2000 (UT), and at the 1.54 Danish
Telescope of La Silla European Southern Observatory, on May 29/30,
2000 (UT). Standard IRAF tasks were used for flat field correction, bias
subtraction, cosmic ray removal, extraction and wavelength-flux
calibration.

We used the Boller \&  Chivens spectrograph installed  in the
Cassegrain  focus of the 2.12\,m  telescope at OAN SPM.  The
300 lines/mm grating was  used to cover a wavelength range
from 3700 to 7600\,\AA. The TEK $1024\times 1024$ pixel CCD with
a $0.24\mu$m pixel size was attached to the spectrograph.

The Danish Faint Object Spectrograph
and Camera (DFOSC) was used with grisms 5 and 7 at the Danish
telescope in Chile, yielding  a combined wavelength range of 3500--9500
\AA. A slit width of 1.5 arcseconds was chosen in order to match the
typical point spread function at the focal plane of the telescope.
This resulted in spectral resolutions of 4 \AA~(grism 7) and 7.5 \AA~(grism
5).
The journal of observations is given in
Table 1.

In general, the wavelength calibration functions were constructed with
$\approx$ 40 He-Ne lines and had typical $rms$ of 0.1 \AA~(5 km s$^{-1}$ at
H$\alpha$).

Spectrophotometric  standard stars (Feige\,34 and HZ\,44 in OAN SPM and
LTT\,3864 and LTT\,7987 in La Silla, \cite{hamuy92}, \cite{hamuy94})
were observed   in order  to perform flux calibration.
To minimize slit losses and improve flux
calibration, a wide slit was used for standards stars (5\arcsec) while  the
slit for all objects was aligned along the paralactic angle.
The atmospheric absorption
bands were removed from the spectra using the "telluric" IRAF task. For
that we used a template obtained by normalizing the standard star
spectrum to the continuum, and interpolating the resulting spectrum between
the hydrogen absorption lines intrinsic to the early spectral type.

\begin{table}
\caption[]{Journal of spectroscopic observations. Telescope and grating,
resolution and number of spectra are given. Zero point for Heliocentric
Julian Day is $HJD_{0}$ = 2451600.}  \begin{center}
\begin{tabular}{lcccc} \hline  \multicolumn{1}{c}{Tel.}&
\multicolumn{1}{c}{Grating} & \multicolumn{1}{c}{Res. (\AA)} &
\multicolumn{1}{c}{HJD}& \multicolumn{1}{c}{N} \\
\hline  2.12m/SPM &300
l/mm  &8.0 &45.6676-7395& 11\\ 1.54m/ESO &\#5       &7.5 &94.5003-5759& 6
\\ 1.54m/ESO &\#7 &4.0 &95.4913-5675& 17\\   \hline \end{tabular}
\end{center} \end{table}

 \begin{figure}
  \includegraphics[angle=0,width=9cm]{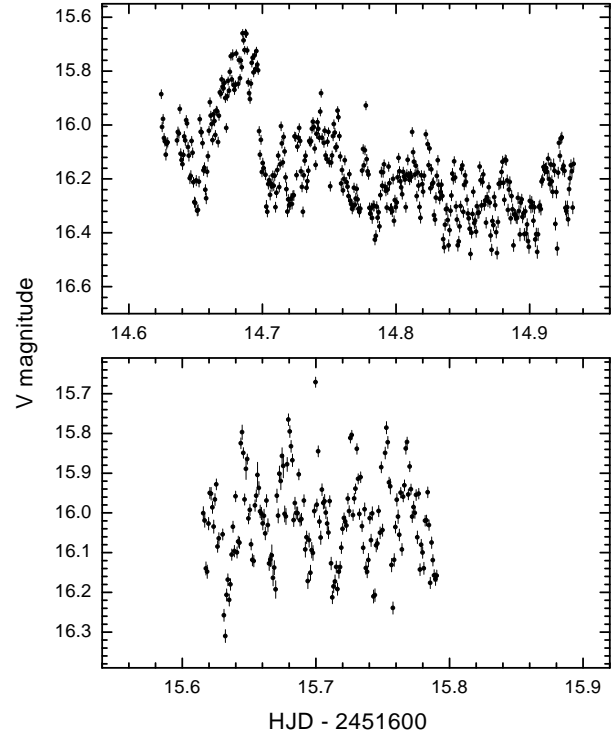}
  \caption{V magnitudes and their errors for the observations at Braeside
Observatory.} \label{spectrum} \end{figure}

        \begin{figure}
  \includegraphics[angle=0,width=9cm]{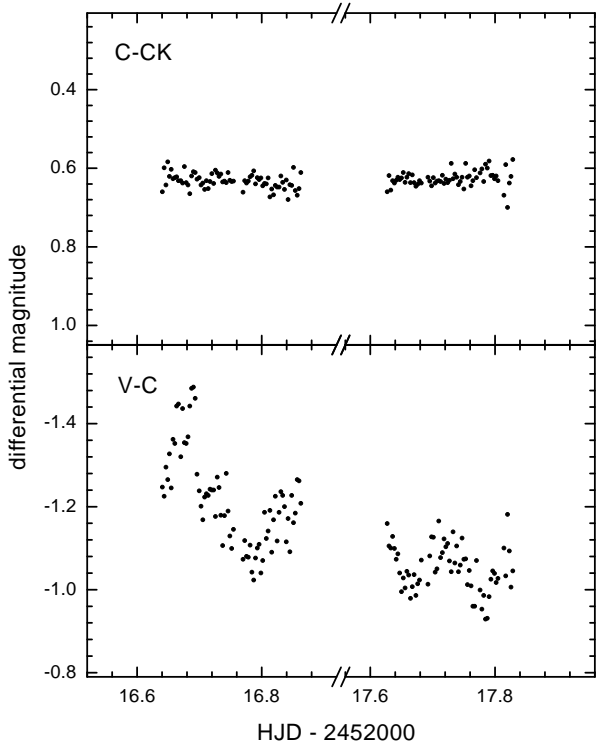}
  \caption{Differential magnitudes of comparison and check star (upper
panel) and variable and comparison star (bottom
panel) for the San Pedro Martir Observatory observations.}
\label{spectrum} \end{figure}

    \begin{figure}
  \includegraphics[angle=0,width=9cm]{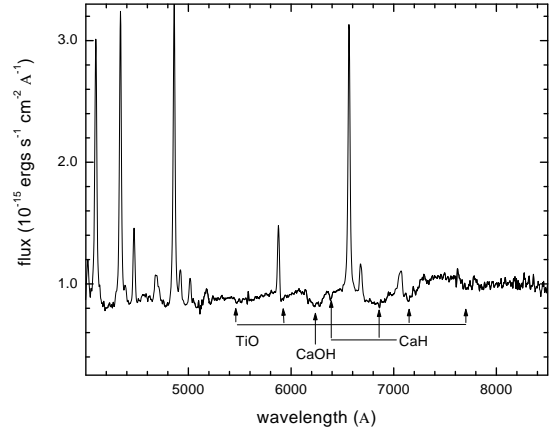}
  \caption{Combined spectrum of \object{1RXS\,J094432.1+035738}.
Identified
absorption bands are labeled. }
  \label{spectrum} \end{figure}

 \begin{table}
\caption[]{Spectral lines detected in the spectrum of
\object{1RXS\,J094432.1+035738}. We give the equivalent width in \AA~ and
the flux  in 10$^{-15}$ erg s$^{-1}$ cm$^{-2}$ \AA$^{-1}$.} \begin{center}
\begin{tabular}{lrc} \hline  \multicolumn{1}{c}{Line}&
\multicolumn{1}{c}{EW} & \multicolumn{1}{c}{flux}  \\ \hline
H10  &N/A &2.1\\
 H9/\ion{He}{I} 3828 &-9 &2.2\\
  H8 &-20 &2.4\\
 H$\epsilon$ 3963 &-37 &2.7\\
 \ion{He}{I} 4023 &-5 &1.1\\
  H$\delta$ &-65 &2.8\\
  H$\gamma$  &-70 &3.0\\
   \ion{He}{I} 4388 &-4 &0.9\\
  \ion{He}{I} 4471 &-15 &1.4\\
 \ion{He}{II} 4685 / \ion{He}{I} 4713 &-11 &1.0\\
 H$\beta$ & -72 &3.1  \\
 \ion{He}{I} 4920 &-8 &1.0\\
 \ion{He}{I} 5015 &-4 & 1.0\\
 \ion{He}{I} 5875 &-14 &1.5\\
 H$\alpha$ &-65 & 3.1\\
 \ion{He}{I} 6678 &-8 & 1.2\\
 \ion{He}{I} 7065 &-8 &1.1\\
 \ion{H}{I} 8863 &-5 &N/A \\
  \ion{H}{I} 9014 &-9 &N/A\\
  \ion{H}{I} 9229 &-13  &N/A  \\
 \hline  \end{tabular}
\end{center} \end{table}
\vskip 0.3cm
\normalsize

\begin{figure}
 \includegraphics[angle=0,width=8cm]{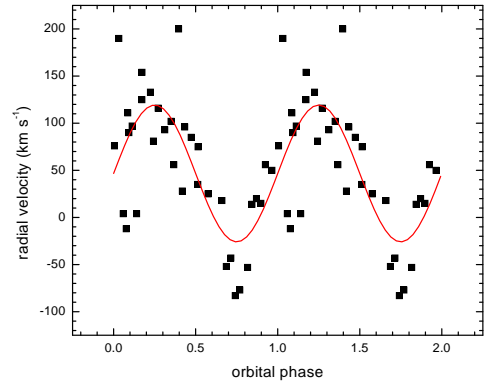}
\caption{The H$\alpha$ emission line radial velocities folded with the 0.14892 d period. The
best sinus fit, with half-amplitude 73 $\pm$ 8 km s$^{-1}$ and zero point
47 $\pm$ 6 km s$^{-1}$, is also shown. The time for radial velocity
croosing from blue to red coresponds to HJD = 2\,451\,645.6538(30).}
\label{spectrum} \end{figure}


\begin{figure}
\includegraphics[angle=0,width=9cm]{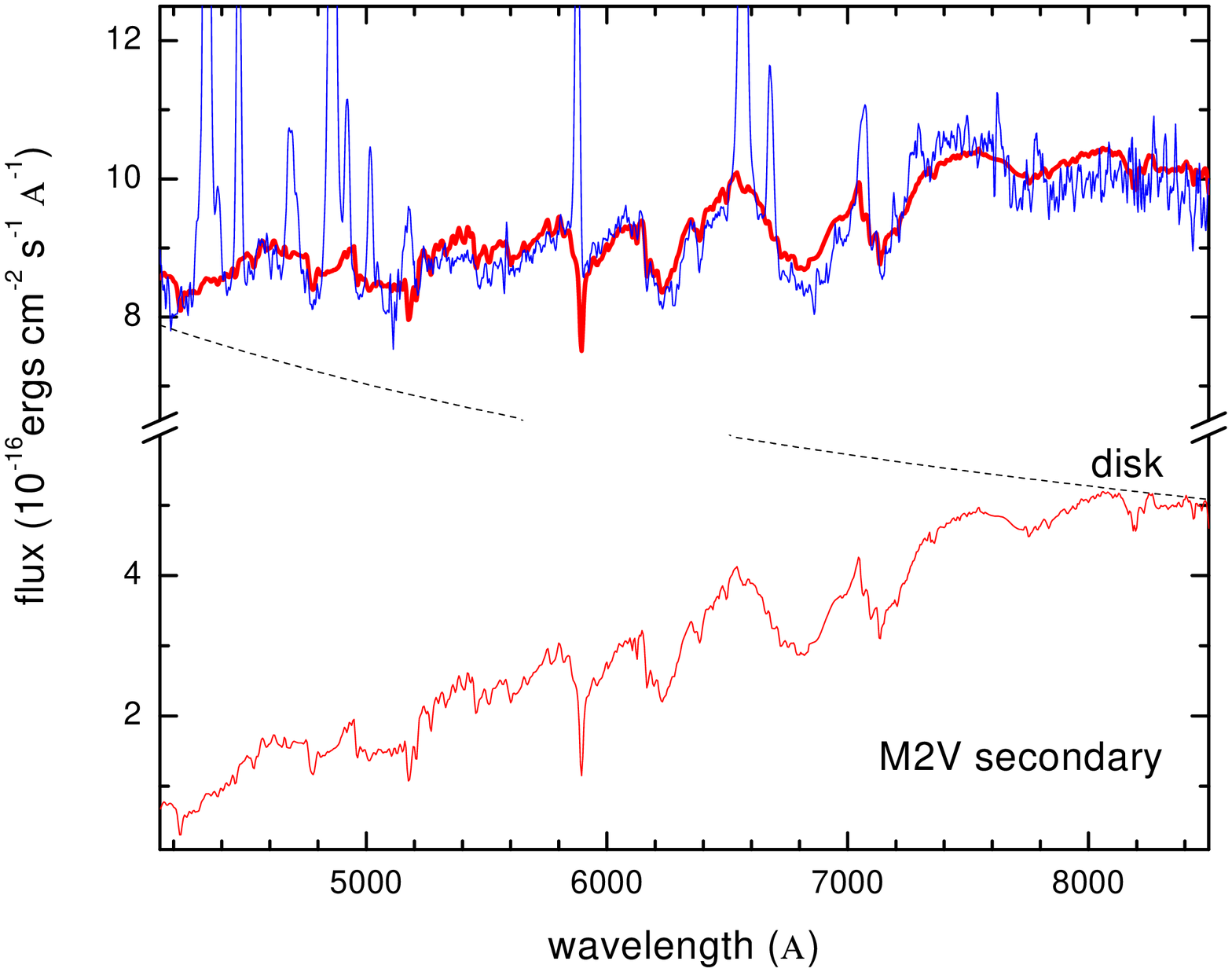}
\caption{The observed spectrum, the best disk and secondary star model
and the sum of them (thick line). Spectral regions with emission lines were
excluded in the fitting process. }   \label{spectrum} \end{figure}

\section{Results}

\subsection{Photometry}

Figs.\, 1 and 2 show the light curves of \object{1RXS\,J094432.1+035738}.
From these figures,  the remarkable variability of this star is evident.
The upper panel of Fig.\,1 reveals
non-coherent variability with amplitude 0.4 mag and peak-to-peak
time scales of 1 hour, whereas the following night a complex behaviour is
seen; a long-term oscillation with a time scale of 1.7 hours over a
long-term fading  and flickering in time scales longer than 10 minutes.
Regarding the data taken in San Pedro Martir Observatory (Fig.\,2),
discarding the data previous to the maximum at 2\,452\,016.6905, and
shifting the remaining data to a common nightly mean, we find a good model
with a sinusoid with a period of 0.111 $\pm$ 0.005 days (2.66 $\pm$ 0.12
h). A frequency analysis of the whole dataset did not yield any conclusive
periodicity.

\subsection{Spectrum description and radial velocities}

Our averaged spectrum shown in Fig.\,3 reveals a rather flat continuum
reaching a maximum around 7500 \AA~ and decreasing to longer wavelengths.
Balmer, Paschen (not shown in Fig.\,3 since the flux calibration at these
wavelengths is not reliable) and \ion{He}{I}  emission lines are also
present. At our resolution, these lines are single peaked and the Balmer
decrement rather flat. From the spectrum we derived spectrophotometric
magnitudes of V = 16.7 $\pm$ 0.1 for April 2000 and V = 16.4 $\pm$ 0.1 for
May 2000, including correction factors by slit looses derived for  the
finite slit width. The above figures indicate that the object  was
observed in quiescence in both observing sessions. Contrasting with the
spectrum shown by  Jiang et al.\,(2000), we observe the TiO absorption
bands typical of M type stars.

We give in Table 1 the fluxes and equivalent
widths for the main emission lines detected in the averaged spectrum.
Typical errors are of the order of 10\%. The average of
the full width at half intensity for hydrogen lines is 1580 $\pm$ 235
km s$^{-1}$ and for neutral helium lines is 1340 $\pm$ 270 km s$^{-1}$.
The mean H$\alpha$ equivalent width was -59 $\pm$ 3 \AA~ on April 2000,
and -50 $\pm$ 4 \AA~ on May 2000. The corresponding figures for
\ion{He}{I} 5875 were -15 $\pm$ 1 \AA~ and -14 $\pm$ 2 \AA, respectively.
We compared the observed Balmer decrements with theoretical results by
Williams (1991), who modeled optically thin gas in the emission lines in
accretion disks. He computed H$\beta$ strengths and Balmer decrements for
a grid of temperatures,  inclinations and mid-plane accretion disk density
($N_{0}$ in units of nucleons per cm$^{3}$). We find that the observed
$D(H\alpha/H\beta$) = 1.35 and   $D(H\gamma/H\beta$) = 0.86 nicely fit the
results for the T = 15.000 $K$, $\log{N}$ = 12.5 model.

We measured the
radial velocities of the H$\alpha$ emission line in the single spectra by
fitting the profile with a Gauss function.
We observe a radial
velocity maximum at HJD 2\,451\,645.6872 (April 12, 2000) followed by a
smooth velocity decrease. On the other hand, on May 23,
2000, we observe a velocity minimum at HJD
2\,451\,695.4992 followed by a maximum at  HJD 2\,451\,695.5586.
A pdm periodogram for our radial velocity
dataset indicates several  possible solutions.
Our velocities are not distributed in time in such way as to decide
about these possible aliases. However, the data provided by Thorstensen
\& Fenton (private comunication) is well distributed in time to determine
the daily cycle count. We therefore assumed the Thorstensen \& Fenton period (0.1492
$\pm$ 0.0004 d) and tried to
refine it by doing a pdm analysis inside the period error window provided by these
authors. The result indicates two solutions with similar statistical
significance, viz.\,
0.14892 $\pm$ 0.00013 d and 0.14936 $\pm$ 0.00013 d.
The period error above
corresponds to the half width at half maximum of the main peak in the
power window. In both cases, the H$\alpha$ radial velocity curve
can be described by a sine law with half-amplitude around  75
km s$^{-1}$ (Fig.\,5).
The fact that $K_{1}$ is moderately high, and the light curve does
shows variations, but no eclipses, point to a medium inclination in the
range of 30--60 degrees.


\subsection{Modeling the spectral energy distribution: the secondary star
revealed}

The existence of the TiO band at 5450 \AA~ sets a
low limit for the spectral type of the secondary star of M\,0.5, whereas
the absence of the TiO band at 8400 \AA~ sets an upper limit of M\,5 (e.g.
Reid \& Hawley 2000). In order to better constrain the spectral type of
the secondary star, we fitted the continuum with a composite spectrum
consisting of the contributions of a power-law continuum and a late type
template spectrum:\\

$ S(\lambda) = a \times T(\lambda) + b \times \lambda^{c}$ \hfill(2)\\

\noindent
where S is the observed spectrum, T the red dwarf template spectrum,
$\lambda$ the wavelength in angstroms
and a,b,c parameters to be
found. Of course, a and b are constrained to the positive domain.
The data sampling for fitting was selected
avoiding emission lines. A sequence of template spectral types between M0V
and M4V were obtained from Pickles (1998), which are  available on-line at
the CDS (Centre de Données astronomiques de Strasbourg) database
(http://cdsweb.u-strasbg.fr).
We run a Levenberg-Marquardt non-linear least squares fitter
getting the a,b,c  parameters and the $\chi^{2}$ for every template.
Our results
indicated that the best fit is reached with a template of spectral
type M2V and a power law with slope c = -0.61(2). The template with a
spectral type of M0V gives  $\chi^{2}$ larger by a factor 2.2,
whereas the templates with spectral types M1V, M3V and M4V give $\chi^{2}$
larger than the M2V template by 5\%, 25\% and
30\%, respectively.
For the M2V model, shown in Fig.\,5,  the
disk contributes 64\% and the secondary star 36\%
to the total light at $\lambda$ 6700 \AA.

\subsection{Distance estimates}

From the inferred Johnson-V magnitude of the secondary star,
viz.\, 18.04, we estimate a distance of 425 $^{+63}_{-47}$ pc,
assuming $M_{V}$ = 9.9 for the M2V secondary and an
uncertainty in the spectral class of $^{+0.5}_{-1.0}$ subtypes.
An independent estimate
can be obtained using the $M_{V}$(outburst peak) vs. $P_{o}$
relationship calibrated by Warner (1995, his equation 3.4).
Assuming a magnitude at maximum of $V$ = 13.1, based on  the
$VSNET$ reports, we obtain a distance of $d$ = 454 pc. Finally,
using the Beuermann \& Weichhold (1999) method to derive CV distances,
which is based on the flux deficiency $f_{TiO}$ at wavelengths 7500 and
7165 \AA, i.e.\, independent of any disk contamination, we find
distances of 399 $\pm$ 96 pc, 431 $\pm$ 104 pc, and 427 $\pm$ 103
pc for M1V, M2V and M2.5V type secondaries, yielding a mean of 419
pc. In the above calculation  we assumed a radius for the
secondary star of 0.36 $\pm$ 0.09 solar radii, accordingly to the orbital
period of 3.58 h (Smith \& Dhillon 1998, Eq. 11). The average
of the three above estimates is $d$ = 433 pc {\bf with a likely
uncertainty of 100 pc.}

\subsection{X-ray data}

Covered
in the ROSAT survey for 243 s with a mean count-rate
of 0.089 counts/s, the hardness ratio suggests a hard spectrum,
possibly absorbed. Using a thermal bremsstrahlung model with a
fixed temperature of 20 keV, we derive an unabsorbed
X-ray luminosity of 8.8 $\times 10^{29}$ ($d$/100pc)$^{2}$ erg s$^{-1}$
in the 0.1-2.4 keV ROSAT range. Using our distance estimate
above, this figure corresponds to 1.65 $\times$ 10$^{31}$ erg s$^{-1}$,
a typical value for dwarf nova in quiescence (Cordova \& Mason 1984).

\section{Discussion: on the nature of \object{1RXS\,J094432.1+035738} }

The orbital period found by Thorstensen \& Fenton (private
comunication) is quite short for a dwarf nova above the period gap
(Shafter 1992).
However, the low
\ion{He}{II}\,4686/H$\beta$  emission line ratio, along with the
absence of cyclotron harmonics in the spectrum, probably rule-out
the hypothesis of a AM Her type system, which is also consistent with
the available X-ray data.   On the
other hand, the presence of outbursts and the general spectrum appearance,
suggests a dwarf nova type object.  The short-term photometric variability
observed in Fig.\,1 is reminiscent of rotational modulations seen in
intermediate polars, but the changing nature of this feature, and their
non-coherent nature, makes this interpretation doubtful.
The outburst amplitude, $\sim$ 3.5
mag, implies a recurrence time of around 30 days, accordingly to the
Kukarkin-Parenago relationship (e.g.\, Warner 1995, Eq.\,3.1). The above
figure suggests that many outbursts might have been missed by observers in
previous campaigns (the two reported outburst are separated by 6 months).
We argue that \object{1RXS\,J094432.1+035738} is very likely
a U Geminorum type dwarf nova, although an
intermediate polar nature cannot be discarded.

\begin{acknowledgements}
We acknowledge the referee, Dr. Klaus Beuermann,
for useful comments that helped to improve
a first version of this paper. We also
acknowledge Dr. John Thorstensen for his cooperation in
discussing the orbital period of this object.
This work was
supported by Grant Fondecyt 1000324 and DI UdeC 99.11.28-1.
 
\end{acknowledgements}

\end{document}